\documentclass[useAMS,usenatbib]{mn2e}
\usepackage[totalwidth=515pt,totalheight=680pt,left=1.4cm,right=1.4cm]{geometry}
\usepackage{graphicx,amssymb}
\input psfig

\def\OmegaG{{\Omega_{\rm G}}}
\def\Msun{{M_{\odot}}}
\def\br{{\bmath r}}
\def\expr{{\bmath e}_{x'}}
\def\ex{{\bmath e}_{x}}
\def\eypr{{\bmath e}_{y'}}
\def\ey{{\bmath e}_{y}}  
\def\ezpr{{\bmath e}_{z'}}  
\def\ez{{\bmath e}_{z}}  
\def\Rd{{R_{\rm d}}}  
\def\tsurv{{t_{\rm surv}}}  
  
\title[Exoplanets Beyond the Solar Neighbourhood]  
{Exoplanets Beyond the Solar Neighbourhood: Galactic Tidal Perturbations}  
\author[Veras \& Evans]{Dimitri Veras$^{1}$\thanks{E-mail:  
veras@ast.cam.ac.uk}, N. Wyn Evans$^{1}$\thanks{E-mail:  
nwe@ast.cam.ac.uk}\\  
$^{1}$Institute of Astronomy, University of Cambridge, Madingley Road, Cambridge CB3 0HA}  
  
\begin{document}

\date{Accepted 2012 December 16.  Received 2012 December 14; in original form 2012 October 28}  
  
\pagerange{\pageref{firstpage}--\pageref{lastpage}} \pubyear{XXXX}   
  
\maketitle  
  
\label{firstpage}  
  
\begin{abstract}  
The majority of Milky Way extrasolar planets likely reside within a
few kpc of the Galactic centre. The Galactic tidal forces acting on
planets scale inversely with radius in the Galaxy and so are much
greater in the inner Galaxy than in the Solar neighbourhood. Within a
range of 3.5 to 10 kpc, the vertical tide from the Galactic disc is
predominant.  Interior to 3.5 kpc, the effects of the Galactic bulge
cannot be neglected and the in-plane tidal components are as important
as the vertical ones. Here, we quantify the orbital changes induced by
these tides.  We find that the greatest perturbations occur when the
planetary orbit is severely misaligned to the parent star's orbit.
When both planes are perpendicular, the eccentricity of the planet is
driven to unity, although the semimajor axis is secularly unaffected.
When both planes are coincident, the effect from Galactic tides is
minimized, but remains non-zero.  In these cases, we provide estimates
for the survival times, as well as the minimum baseline eccentricity
variation for all Milky Way exoplanets as a function of Galactic
parameters.  Inclinations similar to the Solar System's ($\approx
60^{\circ}$) can easily cause eccentric Neptunes (at $\approx 30$ AU)
around host stars deep within the Galactic bulge (within $50$ pc) to
experience eccentricity variations of several tenths, and cause the
exoplanets with the widest-known separations ($\approx 10^3$ AU) to
experience similar variations in the Galactic disc. These variations
occur on timescales of a few Gyr, a fraction of a typical main
sequence lifetime.
\end{abstract}  
  
\begin{keywords}  
planets and satellites: dynamical evolution and stability --  
planet-star interactions -- fundamental parameters; The Galaxy:  
kinematics and dynamics -- structure -- disc  
\end{keywords}  
  
\section{Introduction}  
  
The vast majority of the thousands of candidate and confirmed  
exoplanets reside in the Solar neighborhood  
\footnote{See the Extrasolar Planet Encyclopedia at  
  http://exoplanet.eu/}$^{,}$\footnote{See the Exoplanet Data Explorer  
  at http://exoplanets.org/}, which is $\sim 8$ kpc from the Galactic  
centre.  Given their abundance locally, it is natural to conclude that  
the whole Galaxy is teeming with exoplanets, in accord with the  
Copernican Principle. Some confirmation is provided by microlensing  
surveys, which typically monitor source stars in the Galactic bulge.  
Intervening host stars can act as gravitational lenses, whilst their  
associated exoplanets can be detectable as perturbations of the  
microlensing lightcurves \citep[e.g.][]{Do09,Ja10,Mi11,Ye12}. In this  
way, exoplanets have been discovered at Galactocentric radii from  
$\sim 3$ to $6$ kpc, as exemplified by OGLE 2007-BLG-050~\citep{Ba09}  
and OGLE-2003-BLG-235~\citep{Be06}. Additionally, the SWEEPS  
(Sagittarius Window Eclipsing Extrasolar Planet Search) collaboration  
identified 15 transiting exoplanet candidates in the Galactic bulge,  
and concluded planets are just as common there as in the Solar  
neighbourhood~\citep{Sa06}.  
  
The density of the Galactic disc increases moving towards the centre  
in a roughly exponential manner with a scale length of between 2 and 3  
kpc~\citep[see e.g.,][]{Bi98}. Additionally, the innermost parts of  
the Milky Way are dominated by a $10^{10} \Msun$ boxy-shaped bulge, as  
seen most prominently in the COBE/DIRBE near-infrared light  
distributions~\citep{Dw95}. As the star density increases  
substantially towards the centre, there are many more possible  
exoplanet hosts in the inner Galaxy than in the remoter outskirts,  
like the Solar neighbourhood. However, the Galactic environment itself  
becomes more extreme, as stellar collisions, encounters, and flybys  
are more frequent in the inner parts and the Galactic tides are  
stronger.  
  
The largest component of the Galactic tidal field in the Galactic disc  
acts perpendicular to the Galactic plane. Unfortunately, observations  
have not yet been able to identify a typical planetary orbit  
inclination with respect to the Galactic plane due to a strong bias:  
the majority of all exoplanetary candidates have been discovered by  
the {\it Kepler} mission \citep{Bo11a,Bo11b,Ba12}, which observes a  
fixed patch of sky along the Orion arm and can detect candidates with  
only nearly edge-on orbits.  Programs to detect transiting planets see  
them edge-on at a wide variety of declinations and right ascensions,  
depending on the instruments' line of sights (see  
\citealt*{vermoe2012} for a more detailed discussion).  The radial  
velocity technique, which is responsible for the discovery of the  
majority of confirmed exoplanets, provides no information about the  
planetary orbital inclination with respect to the Galactic centre.  
Further, the Solar System's invariable plane and ecliptic are  
misaligned with the Galactic plane at an angle of approximately  
$60^{\circ}$ \citep[e.g.][]{Hu66,Du87}.  Although the Galactic  
inclination distribution of exoplanets is unconstrained, high  
inclinations certainly exist, and may even be typical.  
  
Regardless of their orientation and location in the Galaxy, all Milky  
Way exoplanetary systems experience the Galactic tide. Within its Hill  
or Roche surface, a star's gravity dominates and an exoplanet may  
survive the effects of Galactic tides  
unscathed~\citep[e.g.,][]{He86}. Exterior to the Hill surface, the  
Galactic tides are always important. For a $1M_\odot$ star in the  
Solar neighbourhood, the Hill surface has an extent $\sim 10^5$ AU,  
which is a measure of the size of any exoplanetary system. 
At a Galactocentric radius of 500 pc, the Hill surface is an order of  
magnitude smaller with a typical extent of $\sim 10^4$ AU.  For the  
planets in the Solar system, the Galactic tide is not generally  
important.  Although effects of tides on planets in the Solar neighbourhood
has been considered before, studies of tides on exoplanetary
systems in the bulge has been restricted to Oort clouds \citep{Br10}.

A rough rule-of-thumb is that the precession timescale due  
to tides $P_{\rm tide} \approx P_{\rm ext}^2/P_{\rm pl}$, where  
$P_{\rm ext}$ is the orbital period of the host star in the Galaxy and  
$P_{\rm pl}$ is the orbital period of the planet around the star. For  
Jupiter, this gives $P_{\rm tide} \approx 10^{14}$ yr, well in excess  
of a Hubble time. However, wide-separation planets are now  
known~\citep[see e.g.,][]{Go10,Lu11, Ku11} with semimajor axes up to $\sim  
2500$ AU. For such wide-separation planets in the inner Galaxy,  
$P_{\rm tide} \approx 10^{10}$ yr or less. In other words, there can  
be significant effects from Galactic tides over the age of   
such exoplanetary systems.

\subsection{Wide Orbit Planet Motivation}

Given the observationally-motivated emphasis in this work on 
wide-orbit planets, here we 
provide an updated summary of 
\cite{veretal2009}, which describes the prospects for the 
formation, survival and detection of these bodies.

\subsubsection{Formation of Wide-Orbit Planets}

Here we briefly review mechanisms for generating planets at different distances
from their host stars.  The core accretion formation mechanism
\citep[e.g.][]{poletal1996} can readily form planets at several AU but has 
difficulty forming planets at tens of AU.
Even Uranus and Neptune, at $\approx 19$ AU and $\approx 30$ AU, require particularly favorable 
circumstances to have been formed {\it in situ} from core accretion 
\citep{levste2001,thoetal2002}.  \cite{dodetal2009} claim that $35$ AU
is a rough limit beyond which massive gas giant planets must form by an
alternate mechanism, such as gravitational instability in the disc 
\citep[e.g.][]{cameron1978,boss1997}; \cite{boley2009}
claims that this limit is approximately $100$ AU.  Disc instability easily forms 
planets at tens of AU \citep{boss2003,boss2011},
and may or may not be able to form planets at a couple hundred AU \citep{boss2006,boley2009}.

Beyond a few hundred AU, planets are highly unlikely to have been formed {\it in situ}.
Instead, they were likely formed and subsequently scattered outward within the same system due
to multi-planet gravitational instability \citep[e.g.][]{rasfor1996,weimar1996,linida1997}.
\cite{schmen2009} and \cite{veretal2009} showed that planet-planet scattering amongst massive gas giants which could
have been formed by core accretion ($a \lesssim 35$ AU) may generate a population of 
planets from $10^2$ AU - $10^5$ AU.  However, those planets with the widest orbits are 
unlikely to remain bound for several
tens of Myr if they interact with any surviving tight-orbit planets.  \cite{boletal2012} 
performed multi-planet scattering studies assuming one planet was formed by gravitational
instability at $a = 100$ AU.  This initial wide-orbit planet induced significant radial mixing,
ultimately producing a population of planets with separations between several hundred and 
several thousand AU (see their Fig. 15).  Alternatively, wide-orbit planets could 
represent captured free-floaters \citep{perkou2012,varetal2012}, a 
possibility that will become increasingly plausible with additional studies that support
the purported vast free-floating planet population \citep{sumetal2011}.

\subsubsection{Survival of Wide-Orbit Planets}

Planets scattered to stable wide orbits could remain in those orbits if they remain unperturbed.  
One potential destabilizing perturbative source is a remnant inner planet which survives the 
scattering phase.  If the wide-orbit planet is sufficiently eccentric, then repeated 
perturbations by the inner planet will cause an additional scattering event and instability.  
This additional scattering event may not take place for tens of Myr \citep{veretal2009}. 

Another perturbative source is the birth cluster itself.  Although stellar flybys can actually 
aid the passage of a planet onto a wide orbit \citep[e.g.][]{maletal2011,boletal2012}, 
alternatively flybys could also cause instability and ejection.  \cite{adaetal2006} estimate 
that over 10 Myr, for clusters with between 100-1000 members, the typical impact parameter 
for two stars passing each other is about 700-4000 AU.  Any planets at or beyond these distances 
are likely to be severely disrupted.  Whether this disruption triggers ejection or just 
a shift to a different wide orbit is dependent on the geometry of the encounter 
\citep[see, e.g.][]{vermoe2012}.  Planets with tighter orbits also may experience 
instabilities with a wide range of outcomes;  recently \cite{parqua2012} performed 10 Myr 
cluster simulations with single planets on circular orbits at 5 and 30 AU, and found 
escape, orbit disruption, or quiescent evolution were all possible outcomes for individual
stellar systems.

Overall then planetary interaction in clusters is highly model-dependent and disrupted 
planets may or may not survive cluster evolution on wide orbits.  A simpler argument for 
how planets can survive their birth cluster evolution is that in many cases the dissociation 
timescale for the cluster is much shorter than the timescale for a planet to experience a 
strong perturbation from a stellar flyby, or even fully form.  Although the cluster 
dissociation timescale is a function of both gas dynamics as well as N-body dynamics 
\citep[e.g.][]{moeetal2012}, the dissociation timescale appears to be positively correlated 
with the cluster membership population \cite[e.g.][and references therein]{stedur1995,bate2012}.
Therefore, the smaller the cluster, the more likely planets can survive the cluster phase undisturbed.

\subsubsection{Detection of Wide-Orbit Planets}

Although the majority of exoplanets have been discovered with Doppler radial velocity variations 
or transit measurements, these techniques fail to probe the outer reaches of planetary systems.  
The widest-orbit planets are instead discovered by direct high-contrast imaging.  These detections 
often require follow-up imaging studies at different wavelengths and/or with different instruments 
to ensure that the star and companion are kinematically associated with each other (have the same 
proper motion) and to constrain the companion's mass.  For example, the exoplanet with the 
widest-known orbit (a projected separation of $\approx 2500$ AU), WD 0806-661B b, was initially detected by 
\cite{Lu11} but was later confirmed by \cite{luhetal2012}.

All 16 companions at separations of at least 100 AU from their parent
stars\footnote{Numbers are correct as of 18 November 2012 from the
  Extrasolar Planets Encyclopedia.} have super-Jovian masses, and at
least half of these are likely to be brown dwarfs by mass
(\citealt*{spietal2011} show the planet/brown-dwarf boundary to be
$11-16 M_J$).  For WD 0806-661B b, the companion's mass is reported as
$8 \pm 2 M_J$, safely pinpointing this companion as a planet.
Although the distinction between planetary masses and brown dwarf
masses likely helps to indicate the way the objects are formed, the
resulting difference in motion due to Galactic tides is negligible.

Probing this super-Jovian mass regime is important for understanding the low-mass tail of 
the initial mass function and evaluating the extent of the brown dwarf 
desert \citep{marbut2000,grelin2006,deetal2008,pinetal2012}.  Further, exploring planetary 
systems beyond about $10$ AU is important for comparison with the Solar System, and to 
assess the prevalence of exo-Kuiper belts and exo-Oort clouds \citep[e.g.][]{rayarm2012}.  
The Gemini Planet Imager (GPI), with first light due in January 2013, is dedicated to exploring 
outer regions of exosystems where radial velocity and transit surveys cannot reach.  Forthcoming 
GPI discoveries will raise questions about the formation and fate of wide-orbit companions, and are 
likely to revise the current Galactic-wide exoplanet population estimates 
\citep{sumetal2011,casetal2012}.  Our study helps to preempt this line of inquiry by 
considering the effect of Galactic tides on wide-orbit companions at all distances within the Solar Circle.

\subsection{Plan for Paper}

This paper examines how the Galactic tidal field may drive orbital  
evolution in exoplanets for disc and bulge host stars. We first  
rederive the equations of planetary motion subject to the Galactic  
tidal field in Section \ref{sec:galtides}. We build a new 
three-component model of the Galaxy in Section \ref{sec:galaxy} and  
demonstrate the relative importance of each component in different  
regions. We apply these equations in the regions dominated by the  
Galactic disc (Section \ref{sec:disc}) and Galactic bulge (Section  
\ref{sec:bulge}) before characterising the minimum-possible  
eccentricity variation of exoplanets due to tides throughout the  
entire Milky Way in Section \ref{sec:mine}.  We discuss these  
results in Section \ref{discussion} and conclude in  
Section \ref{conclusion}.

\section{The Galactic Tides} \label{sec:galtides}  
  
\subsection{Derivation}  
  
The disturbances on a planetary orbit can be modeled with the  
perturbed two-body problem. Following \citet{He86}, we use a  
non-rotating, rectangular coordinate system $\br = (x,y,z)$ centered  
on, and orbiting with, the host star. The star is assumed to move on a  
circular orbit of radius $R_0$ about the Galactic centre with  
frequency $\OmegaG$. Then the instantaneous unit vectors in the radial  
and tangential directions are  
\begin{eqnarray}  
\expr &=& \cos (\OmegaG t) \, \ex  + \sin (\OmegaG t) \, \ey \\  
\eypr &=& -\sin (\OmegaG t) \, \ex + \cos (\OmegaG t) \, \ey  
\end{eqnarray}  
The triad ($\ex,\ey,\ez$) has fixed spatial directions while  
($\expr,\eypr,\ezpr$) rotates with the star.  At the origin of both  
primed and unprimed coordinate systems, the gravitational force from  
the Galactic potential $\Phi$ exactly balances the centripetal force  
for circular motion  
\begin{equation}  
\OmegaG^2 R_0\expr - \nabla \Phi =0.  
\end{equation}  
At a general point in the exoplanetary system, the force per unit mass  
is the sum of the forces derived from the two-body problem of star and  
planet, the forces due to the rest of the Galaxy, and the fictitious  
forces caused by our choice of a non-inertial frame:  
\begin{equation}  
F = - {G \mu\over r^3} \expr - \nabla \Phi + \OmegaG^2 R_0 \expr  
\end{equation}  
Now, we Taylor expand the Galactic potential about the host star's  
position to obtain  
\begin{eqnarray}  
\Phi(R,z) &=& \Phi\left (R_0\left(1 + {2x'\over R_0} + {x'^2\over R_0^2} + {y'^2\over R_0^2}\right)^{1/2}, z \right) \nonumber\\  
         &\approx & \Phi\left( R_0 + x' + {y'^2\over 2R_0}, z \right)  
\end{eqnarray}  
This expansion implies further that  
\begin{eqnarray}  
\nabla \Phi &=& \left( {\partial \Phi \over \partial R} + x' {\partial^2 \Phi \over \partial R^2} \right)_{(R_0,0)}\expr  
+ {y'\over R_0}  \left( {\partial \Phi \over \partial R} \right)_{(R_0,0)}\eypr   
\nonumber  
\\  
&+& \left( {\partial \Phi \over \partial z} + x' {\partial^2 \Phi \over \partial z^2}\right)_{(R_0,0)} \ezpr  
+ O(x'^2,y'^2,z'^2)\end{eqnarray}  
If the host star lies in the Galactic plane, then the approximate  
symmetry $z \rightarrow -z$ of the Galactic potential ensures that the  
vertical gradient ${\partial \Phi / \partial z}$ vanishes. It is  
conventional to introduce the Oort constants  
\begin{eqnarray}  
A &=& - \left( {R\over 2}{d\OmegaG \over  dR} \right)_{(R_0,0)}\\  
B &=& -\left( \OmegaG  + {R\over 2}{d\OmegaG \over  dR} \right)_{(R_0,0)}  
\end{eqnarray}  
which gives us  
\begin{eqnarray}  
F &=& - {G \mu\over r^3} \expr + (A-B)(3A+B)x'\expr 
\nonumber
\\ 
& - & (A-B)^2y'\eypr   
- \left( {\partial^2 \Phi \over \partial z^2} \right)_{(R_0,0)} z' \ezpr  
\end{eqnarray}  
Finally, we can use Poisson's equation in cylindrical coordinates to  
relate the local density at the host star to the vertical gradients:  
\begin{equation}  
\left( {\partial^2 \Phi \over \partial z^2} \right)_{(R_0,0)} =   
4\pi G \rho(R_0,0) - 2(B^2-A^2).   
\end{equation}  
We thus obtain an equation derived by \citet{He86},  
\begin{eqnarray}  
F &=& - {G \mu\over r^3} \expr + (A-B)(3A+B)x'\expr 
\nonumber
\\ 
& - & (A-B)^2y'\eypr   
- \left( 4 \pi G \rho -2(B^2-A^2) \right) z' \ezpr  
\end{eqnarray}  
By converting from the primed to the unprimed coordinate system, this  
equation takes the form  
\begin{eqnarray}  
\frac{d^2x}{dt^2} &=& - \frac{G \left(m_{\star} + m_p \right) x}{\left(x^2 + y^2 + z^2\right)^{3/2}}   
+ \Upsilon_{xx} x + \Upsilon_{xy} y  
\label{xeq}  
\\  
\frac{d^2y}{dt^2} &=& - \frac{G \left(m_{\star} + m_p \right) y}{\left(x^2 + y^2 + z^2\right)^{3/2}}   
+ \Upsilon_{yx} x + \Upsilon_{yy} y  
\label{yeq}  
\\  
\frac{d^2z}{dt^2} &=& - \frac{G \left(m_{\star} + m_p \right) z}{\left(x^2 + y^2 + z^2\right)^{3/2}}   
+ \Upsilon_{zz} z   
\label{zeq}  
\end{eqnarray}  
where $m_{\star}$ and $m_p$ represent the masses of the star and  
planet, respectively, and the perturbations $\Upsilon$ are  
\begin{eqnarray}  
\Upsilon_{xx} &=&  \Omega_{G}^2 [ (1-\delta) \cos{\left(2 \Omega_G t\right)} -\delta ]  
\label{xx}  
\\  
\Upsilon_{xy} &=&  \Omega_{G}^2 (1 - \delta) \sin{\left(2 \Omega_G t\right)}  
\label{xy}  
\\  
\Upsilon_{yx} &=&  \Omega_{G}^2 (1 - \delta) \sin{\left(2 \Omega_G t\right)}  
\label{yx}  
\\  
\Upsilon_{yy} &=&  -\Omega_{G}^2 [ (1 -\delta)\cos{\left(2 \Omega_G t\right)} + \delta]  
\label{yy}  
\\  
\Upsilon_{zz} &=&  -\left[ 4 \pi G \rho(R_0,0) -2 \delta \OmegaG^2 \right]  
\label{zz}  
\end{eqnarray}  
where $\delta = - (A-B)/(A+B)$. This form of the equations is given in
\citet{Br10}\footnote{There is a sign error in their expression for
  $\Upsilon_{zz}.$}, although our derivation makes it clear that the
reference frame is not inertial, as they mistakenly claim. For any
Galactic model, the Oort constants $A$ and $B$, together with the
shear $\delta$, are calculable from the Galactic potential and vary
with position in the Galaxy.
  
\subsection{Perturbative Equations of Motion} \label{perturb}  
  
The perturbed two-body problem is usually written down in terms of  
orbital elements, which provide greater intuition into how the size  
and shape of a Keplerian ellipse changes under the perturbations.  We  
can express the equations of motion (\ref{xeq})-(\ref{zeq}) in terms  
of orbital elements by using Gauss' theory or Lagrange's equations  
(\citealt*{Bu76} and pgs. 54-57 of \citealt*{Mu99}).  An alternative  
procedure for deriving the orbital equations subject to very general  
perturbative forces has been given recently by \cite{Ve12}.  
  
A commonly-used approximation in non-linear systems is to separate the  
fast and slow oscillation variables. In celestial mechanics, this procedure  
can be performed by averaging over the fast oscillations of the mean anomaly or  
true anomaly of the orbiting body.  The result is averaged equations for the  
evolution of the more slowly oscillating variables, such as the  
semimajor axis or eccentricity.  This ``adiabatic'' approximation is   
valid when $\Upsilon/n^2 \ll 1$, where $n$ is the mean motion of the planet.    
Effectively, for the Milky Way, this relation holds whenever the   
semimajor axis $a < 10^4$ AU.  
  
In the general case, a planetary orbit is not coplanar with the  
Galactic disc. In the Solar neighbourhood, the out-of-plane tidal  
component $\Upsilon_{zz}$ is an order of magnitude greater than the  
in-plane components. Thus, the vertical tide usually dominates the  
motion, unless the inclination is very low. This approximation has   
been used to simplify the equations in many previous  
studies~\citep{He86,Ma89,Ma92,Ma95,Br96,Br01,Br05}. The vertical   
adiabatic equations of motion are:  
\begin{eqnarray}  
\left(\frac{da}{dt}\right)_v  
&=& 0  
\label{adiva}  
\\  
\left(\frac{de}{dt}\right)_v  
&=&  
-\frac{5e \sqrt{1 - e^2}}{2 n} \cos{\omega} \sin{\omega} \sin^2{i}  
\Upsilon_{zz}   
\label{adive}  
\\  
\left(\frac{di}{dt}\right)_v  
&=&    
\frac{5e^2\sin{2\omega}\sin{2i}}{8n\sqrt{1 - e^2}}  
\Upsilon_{zz}   
\label{adivi}  
\\  
\left(\frac{d\Omega}{dt}\right)_v  
&=&   
\frac{\cos{i} \left(2 + 3 e^2 - 5 e^2 \cos{2 \omega} \right)}{4n\sqrt{1 - e^2}}  
\Upsilon_{zz}  
\label{adOvi}  
\\  
\left(\frac{d\omega}{dt}\right)_v  
&=&  
\frac{5\sin^2{\omega} \left(\sin^2i - e^2\right) - \left(1 - e^2\right)}{2n\sqrt{1-e^2}}  
\Upsilon_{zz}  
\label{adivvarpi}  
\end{eqnarray}  
Planetary orbital elements appearing in these equations are the  
semimajor axis, $a$, eccentricity, $e$, inclination, $i$, longitude of  
ascending node, $\Omega$, and argument of pericentre, $\omega$.  One  
may confirm that these equations reduce to other results found in the  
literature \citep{Br01,Fo04,Fo06}.  Equation (\ref{adiva}) is a   
consequence of the averaged Hamiltonian of the system being  
independent of the mean anomaly or true anomaly.  
  
Although these equations are useful for studies in the Solar  
neighbourhood, the planar tides are important for other  
locations in the Galaxy.  The planar adiabatic equations of motion can  
be derived from the equations in \citet{Ve12} and are written out in  
Appendix A. As tidal forces are derived from a gravitational  
potential, it is always true that $\Upsilon_{xy} =  
\Upsilon_{yx}$. This equality has the consequence that the planet's semimajor  
axis is never secularly affected by any tidal perturbations (see e.g.,  
Eqs.~\ref{adiva} and \ref{adia}).  Consequently, if the  
eccentricity tends towards unity, then concurrently the  
periastron tends towards zero and the apastron tends towards  
$2a$.  In this case, the planet will either collide with the star  
or leave the adiabatic regime.  The latter is likely to  
occur at distances of several $10^4$ AU \citep{Ve12},  
and must occur if the planet is to escape the system.

\section{The Model of the Galaxy}\label{sec:galaxy}  
  
\subsection{Galactic Bulge, Disc and Halo Models}  
  
In all the numerical calculations in this paper, we use a  
three-component Galaxy model that reproduces observed local   
stellar kinematics data. The Galactic halo is represented by a logarithmic  
potential of the form  
\begin{eqnarray}  
  \label{halopot}  
  \Phi_{\rm halo}(R,z) & = & \frac{v_{0}^{2}}{2} \ln \left( R^{2} +  
    z^2 q^{-2} +  d^{2} \right),  
\end{eqnarray}  
with $v_{0}=215$~km\,s$^{-1}$ and $d=16$~kpc (where $R$ and $z$ are  
cylindrical coordinates). The parameter $q$ is the axis ratio of the  
equipotentials, and controls whether the halo is spherical ($q=1$),  
oblate ($q <1$) or prolate ($q >1$).  The Galactic bulge is modelled  
as a Hernquist potential  
\begin{eqnarray}  
  \label{bulgepot}  
  \Phi_{\rm bulge}(r) & = & \frac{G M_{\rm b}} {r+\epsilon},  
\end{eqnarray}  
using $M_{\rm b} = 3.6 \times 10^{10}$~M$_{\odot}$ and  
$\epsilon=0.7$~kpc. The bulge and halo are very similar to those used by  
\citet{Fe06}.  
  
Many authors use Miyamoto-Nagai discs to represent the Galactic disc  
because of the former's simple form of the potential~\citep[see e.g.][]{Fe06, Br10}. We  
have chosen not to do so here.  Instead, we choose a more realistic  
exponential disc at the cost of more complex analytics; the quantities derived  
from the gravitational potential involve special functions.  If the  
Galactic disc is razor-thin and exponential with scalelength $\Rd$, its  
surface density and rotation curve are~\citep[Pg. 77 of][]{Bi87}  
\begin{eqnarray}  
\Sigma(R,z=0) &=& \Sigma_0 \exp  
\left(-\frac{R}{\Rd}\right),\\ v^2_{\rm circ}(R,z=0) &=& {\pi G  
  \Sigma_0R^2 \over \Rd} \Bigl[ I_0\left( \frac{R}{2\Rd} \right)  
  K_0\left( \frac{R}{2\Rd} \right)\nonumber \\ &-& I_1\left( \frac{R}{2\Rd}  
  \right) K_1\left( \frac{R}{2\Rd} \right) \Bigr]  
\label{disc}  
\end{eqnarray}  
where $I_0$, $I_1$, $K_0$ and $K_1$ are modified Bessel functions. The  
scalelength $\Rd$ is 3 kpc, while the normalisation constant  
$\Sigma_0$ is chosen so as to reproduce the local column disc density  
of $51 M_\odot$pc$^{-2}$~\citep{Fl94}.  Although we have only given  
the result in the plane, the full three-dimensional potential is  
known~\citep{Ev92}.  
  
The superposition of these components gives a good representation of
the Milky Way's rotation curve. The circular speed at the solar radius
is $\sim 220$~km\,s$^{-1}$. The Oort constants at the solar radius are
$A = 14.5$ km\,s$^{-1}$ kpc$^{-1}$ and $B = -12.9$ km\,s$^{-1}$
kpc$^{-1}$.  These values are consistent with determinations of the
Oort constants from Hipparcos data~\citep{Fe97}, which yield $A = 14.8
\pm 0.8$ km\,s$^{-1}$ kpc$^{-1}$ and $B = -12.4 \pm 0.6$ km\,s$^{-1}$
kpc$^{-1}$. The three-component model therefore accurately reproduces
all the local stellar kinematics within the quoted error bars. Away
from the Solar neighbourhood, there is considerable uncertainty in the
values of the circular speed and the Oort constants.
  
Lastly, the vertical tides depend on the density in the Galactic  
plane. The density of the bulge and halo can be straightforwardly  
generated through Poisson's equation. To generate the  
three-dimensional density of the disc, we take the column density from  
Eq.~(\ref{disc}) and smear uniformly over the scaleheight, $h$, of the  
thin disc, where $h \approx 300$ pc~\citep[see e.g.,][]{Bi98}.  
Doing so is tantamount to assuming that the vertical disc  
distribution is exponential, for which there is good observational  
evidence.

\begin{figure}  
\centerline{\psfig{figure=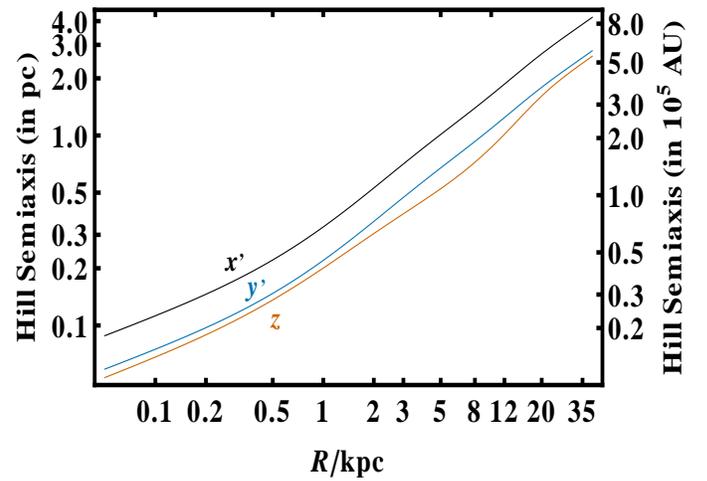,height=7cm,width=9cm} }  
\caption{The semiaxes of the Hill surface in the $x'$ (black), $y'$  
  (blue) and $z$ (orange) directions as a function of Galactocentric  
  radius. Notice that the Hill surface is most elongated in the $x'$  
  direction (along the line to the Galactic centre) and most  
  compressed in the $z$ direction (perpendicular to the Galactic  
  disc).}  
\label{fig:Hill}  
\end{figure}  

The Hill surface in the Galactic tidal field is estimated using the
constancy of the Jacobi integral in \cite{An72}. To a good
approximation, this is an ellipsoidal surface that is stretched along
the line joining the star to the center of the Galaxy, and compressed
in the two orthogonal directions, such that
\begin{equation}  
{x'^2 \over a'^2} + {y'^2 \over b'^2} + {z'^2 \over c'^2}  \approx 1  
\end{equation}  
The semiaxes in the Galactic plane are straightforward to find~\citep{An72}  
\begin{equation}  
a' =  \left(  {G m_\star \over \alpha}\right)^{1/3}, \qquad b' =  
{2\over 3}\left(  {G m_\star \over \alpha}\right)^{1/3},  
\end{equation}  
with $\alpha = 4A(B-A)$. The semiaxis in the direction perpendicular  
to the Galactic plane is more difficult to compute and an explicit  
formula does not seem to have been given before. We find:  
\begin{equation}  
c' = \left( {[Q(1\!+\!\sqrt{1\!+\!Q})]^{2/3}\!-\!Q \over [Q(1\!+\!\sqrt{1\!+\!Q})]^{1/3}}  
\right) \left(  {G m_\star \over \alpha}\right)^{1/3}  
\end{equation}  
where $Q = 4A(A-B)/\Upsilon_{zz}$.  

The variation of the semiaxes of  
the Hill surface with Galactocentric radius are shown in  
Fig.~\ref{fig:Hill}. At the Solar neighbourhood, the surface has  
semiaxes ($1.39, 0.93, 0.72$) pc or ($2.89, 1.92, 1.49$) $\times 10^5$  
AU for a solar mass star.  At 0.5 kpc from the Galactic centre, the  
semiaxes have shrunk by an order of magnitude to ($0.22, 0.15, 0.14$)  
pc or ($4.56, 3.04, 2.81$) $\times 10^4$ AU. The Hill surface is  
important as it gives the typical separation of a planet from its host  
star at which tidal effects become dominant. Within the Hill surface,  
a planet may nonetheless meander because of tidal forces, but its  
wanderings will not usually lead to escape. Note too that the Hill  
surface is much more flattened in the Solar neighbourhood than in the  
Galactic bulge, because the matter distribution is more strongly  
dominated by the Galactic disc at the former location.

\begin{figure}  
\centerline{\psfig{figure=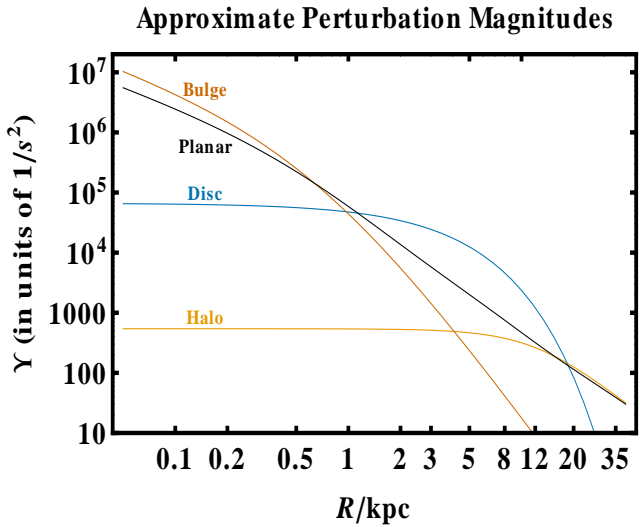,height=7cm,width=9cm} }  
\caption{ A comparison of the amplitudes of the Galactic tidal forces  
  throughout the Milky Way.  The vertical tide is composed of three  
  linearly additive components (bulge, planar and disc) all of which  
  yield no net perturbation when $i = 0^{\circ}$.  The plot  
  demonstrates that the vertical tide is at least 5 times as strong as  
  the planar tide only in the range $3.5$ kpc $\lesssim R \lesssim 10$  
  kpc.}  
\label{fig:comparison}  
\end{figure}  

Although our Galactic model is complicated, it is worth noting a  
helpful rule-of-thumb.  Outside the inner kiloparsecs, the  
three-component model has an almost flat rotation curve with amplitude  
$\approx 220$ kms$^{-1}$. The circular frequency and the Oort  
constants are roughly given by  
\begin{equation}  
A(R)  \approx  -B(R) \approx \frac{110 {\rm\, kms}^{-1}}{R},  
\label{AB}  
\end{equation}  
\begin{equation}  
\OmegaG (R) = A(R) - B(R) \approx    
\frac{220\, {\rm kms}^{-1}}{R}.  
\label{Oort}  
\end{equation}  
Although we always use the full expressions derived from the potential  
of all three components in our numerical calculations, these simple  
expressions are useful in garnering physical intuition.  
  
\subsection{The Galactic Regimes}   
  
Here, we evaluate the contribution of each component in different  
regions of the Milky Way. Doing so helps us to understand the dominant  
effects and to assess the validity of the popular practise of  
neglecting the planar tides often used in previous works  
\citep{He86,Ma89,Ma92,Ma95,Br96,Br01,Br05}.  
  
We summarize the comparison in Fig. \ref{fig:comparison}.  We take $\rho_G  
= \rho_{\rm bulge} + \rho_{\rm disc} + \rho_{\rm halo}$ and derive the  
three curves labeled ``Bulge'', ``Disc'' and ``Halo'' from this  
partition using Eq. (\ref{zz}) with $\delta = 0$.  The black  
``Planar'' curve is traced from $\Upsilon = [\Omega_G(R)]^2$, which  
represents the amplitude of the right-hand side in  
Eqs. (\ref{xx})-(\ref{yy}).  The planar perturbations are a function  
of time, unlike the vertical ones.  
  
Figure \ref{fig:comparison} shows that the planar contribution to the total  
tide can reach at least 10\% of the total in all regions of the  
Galaxy, and may reach $\approx 50\%$ where the bulge and halo are  
important.  Previous studies' neglect of the planar tide is then  
justified at the $\approx 10\%$-level, but only in the regime in which  
the tidal contribution of the Galactic disc dominates. This regime is  
typically over Galacotocentric radii between 3.5 and 10  
kpc. Therefore, we use this approximation too in Section  
\ref{sec:disc} for our study of the disc regime.  Our bulge  
calculations in Section \ref{sec:bulge}, however, all include the  
planar tide, in addition to the vertical tide.

\begin{figure}  
\centerline{\psfig{figure=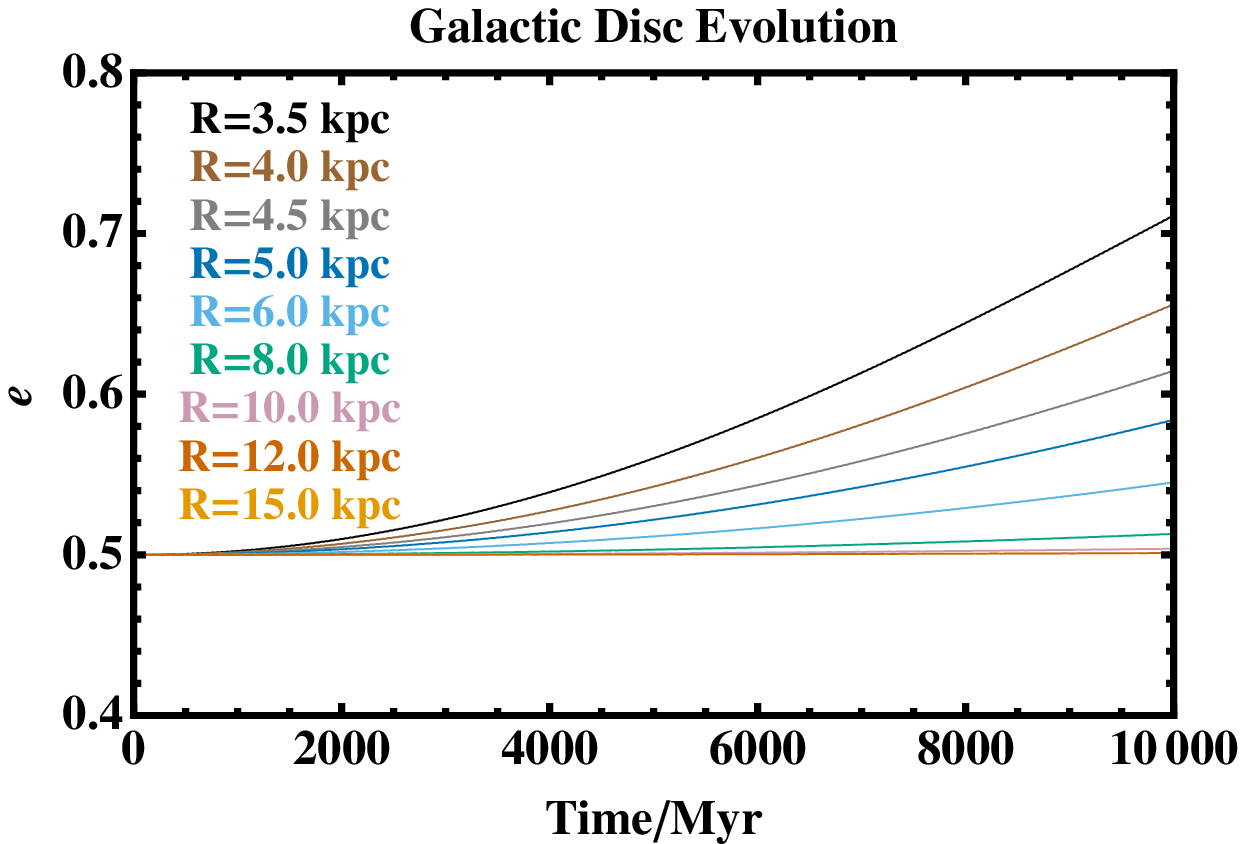,width=9cm}}  
\centerline{\psfig{figure=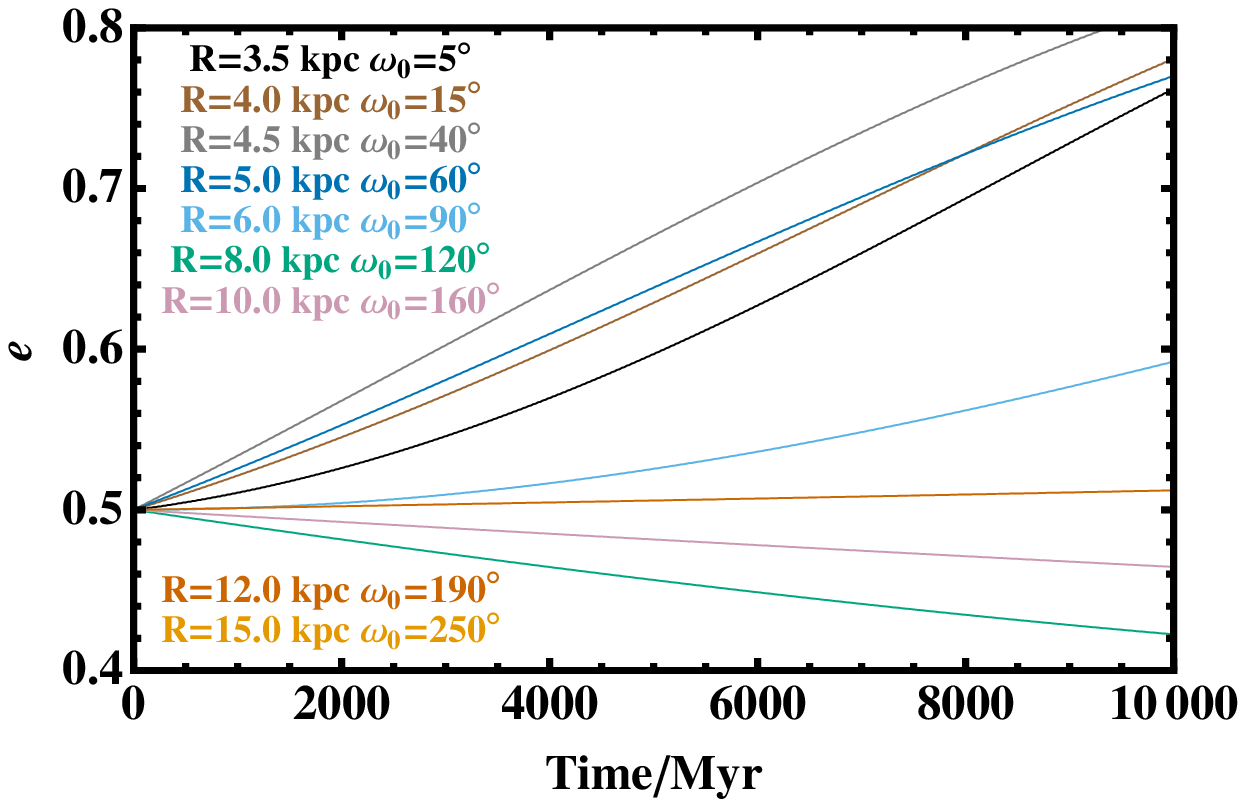,width=9cm}}  
\caption{Planetary eccentricity evolution at different places in the  
  Galactic disc for a Solar System-like inclination ($i = 60^{\circ}$)
  and a wide-orbit planet ($a = 1000$ AU).  In the upper  
  panel, $\varpi_0 = 0^{\circ}$ for each curve.  The lower panel  
  demonstrates that the eccentricity may increase or decrease at  
  different rates depending on the value of $\varpi_0$.}  
\label{fig:discone}  
\end{figure}  

Depending on the accuracy sought, only one vertical component of the Milky  
Way (bulge or disc or halo) needs to be included for many regions of  
the Galaxy.  However, around the bulge-disc transition region, at $R  
\approx $ 1 kpc, and around the disc-halo transition region, at about  
$R \approx 20$ kpc, two components must be included.

\section{The Galactic Disc Regime} \label{sec:disc}  
  
\subsection{Fiducial Evolution}  
  
The range $3.5$ kpc $\lesssim R \lesssim 10$ kpc is the most  
straightforward one in which to explore exoplanet orbital evolution.   
In this regime, we need to account for only the  
contribution from the disc ($\rho_{\rm disc}$) in the vertical tides  
(see Fig. \ref{fig:comparison} and Eqs. \ref{adiva}-\ref{adivvarpi}).  The  
initial inclination of the planet can also be crucial in determining  
the effects of the perturbation.  The vertical tide vanishes for $i =  
0^{\circ}$ as $\Upsilon_{zz}$ is vertically symmetric about the  
Galactic plane.  However, there is little evidence to support  
coplanarity amongst planetary systems and the Galactic disc.

We then consider a planetary system with $i=60^{\circ}$ -- similar  
to the Solar System -- with a one Solar-mass central star.  We assume  
that the star evolves on the main sequence for about 10 Gyr, a value  
which helps motivate the duration of our numerical simulations.  This  
duration is computationally achievable because our simulations are 
adiabatic, and often feature wide-orbit single planets.

\begin{figure*}  
\centerline{  
\psfig{figure=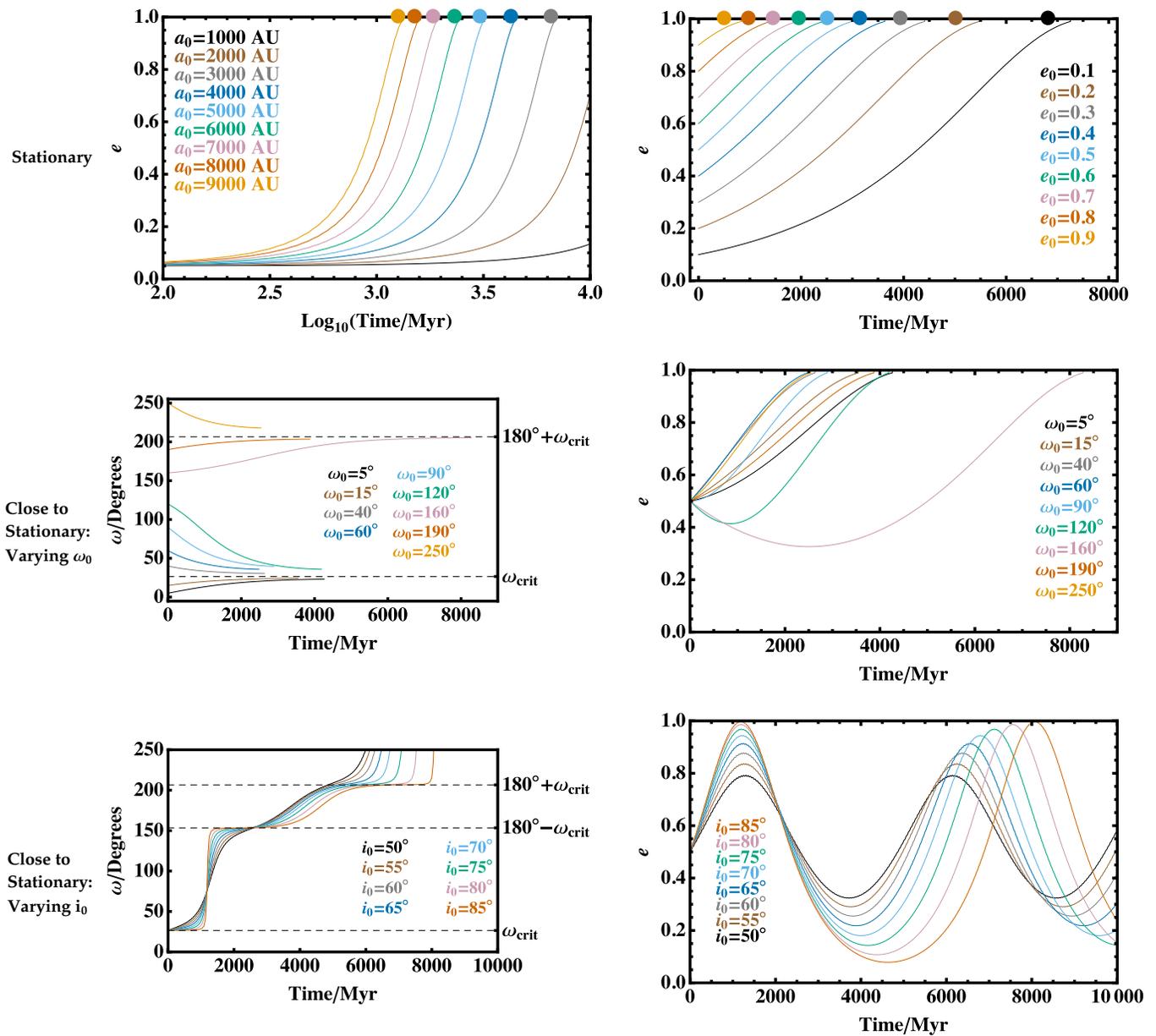,width=18cm}  
}  
\caption{ Characteristics of planetary systems highly inclined with  
  respect to the Galactic plane, a possibly commonplace occurrence.
  For all plots, $R = 4$ kpc.  For the top right plot and middle panels, 
  $a_0 = 2500$ AU.  For the bottom panels, $a_0 = 5000$ AU.
  The upper panel describes planetary evolution arising from polar  
  orbits ($i_0 = i(t) = 90^{\circ}$ or $i_0 = i(t) = 270^{\circ}$) and  
  a stationary argument of pericentre ($\omega_0 = \omega(t) =  
  \omega_{\rm crit}$).  In this regime, given enough time, no planets  
  survive.  Analytical estimates for the survival timescale  
  (Eq. \ref{tsurv}) are shown as large colored dots at $e = 1.0$.  The  
  middle panels describe the motion when $\omega_0$ is allowed to  
  initially deviate from $\omega_{\rm crit}$.  The result is still  
  that no planets survive.  The lower panels keeps $\omega_0$ fixed at  
  $\omega_{\rm crit}$, but allows $i_0$ to deviate from $90^{\circ}$.  
  The result is that planets can persist, although their  
  eccentricities often approach unity.  In the lowest panel, $\Omega_0  
  = 0^{\circ}$ is assumed.  }  
\label{fig:disctwo}  
\end{figure*}  

We provide representative eccentricity evolution profiles for  
wide-orbit planets residing at different locations in the Galactic  
disc in Fig. \ref{fig:discone}.  The upper panel demonstrates the  
dependence on $R$ with a fixed value of $\varpi_0$; the lower panel  
illustrates that changing $\varpi_0$ can have a significant effect on  
the orbital evolution.  Over billions of years, bodies at $a = 10^3$  
AU orbiting stars within the Solar Circle may become significantly  
more or less eccentric.  The smallest values of $R$ typically cause  
greater perturbations, but not always; exoplanetary systems within the  
Solar Circle are likely to harbour more dynamically excited scattered  
discs and Oort clouds than the Solar System.  Microlensing  
observations of wide-orbit eccentric planets residing in the inner parts  
of the disc represent snapshots of dynamically changing systems.

\begin{figure*}  
\centerline{  
\psfig{figure=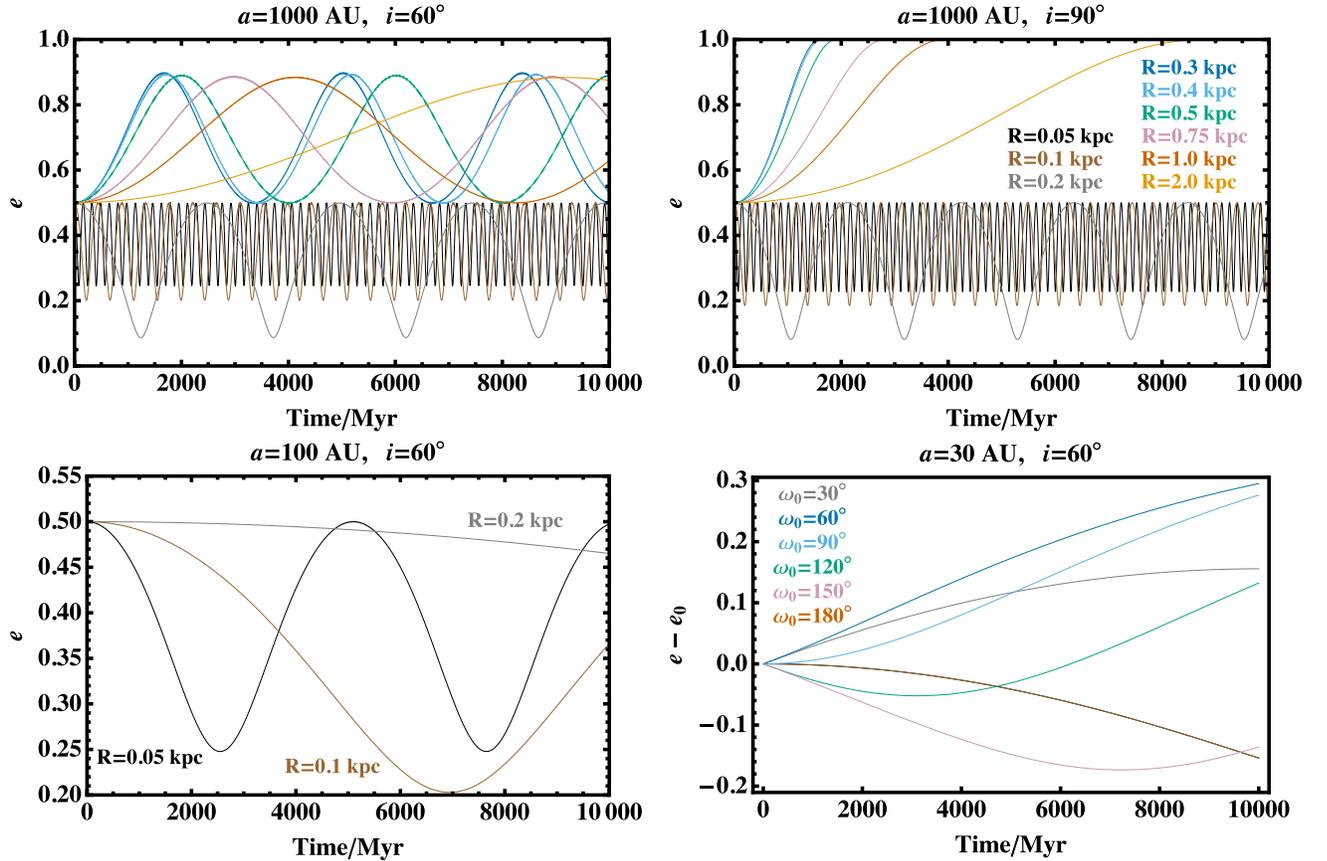,width=18cm}  
}  
\caption{ Planetary eccentricity evolution in the Galactic bulge.  
  Each plot shows the evolution for a different $(a,i)$ pair.  The  
  curves in both upper panel plots are coloured according to the  
  legend in the upper right plot.  All plots assume  
  $\varpi_0=0^{\circ}$, except the lower right plot, for which $R =  
  50$ pc and $e_0 = 0.5$ is assumed for all curves.  This plot  
  demonstrates that eccentric Neptunes deep within the bulge are  
  dynamically active over the course of their main sequence lifetimes.  
}  
\label{fig:bulge1}  
\end{figure*}  

\subsection{Highly-inclined evolution}  
  
As the inclination of a planetary system approaches $90^{\circ}$, the  
variation of the eccentricity is maximized.  At $i = 90^\circ$, the  
inclination does not change, and so polar orbits always remain polar  
orbits. \cite{Br01} found a stationary solution corresponding to  
$\omega_{\rm crit} = \pm\arcsin{(\sqrt{1/5})} \approx \pm  
26.6^{\circ}$; this value may also be deduced from  
Eq. (\ref{adivvarpi}).  The importance of this solution is that,  
whilst $a$, $\Omega$, $i$ and $\omega$ do not change, the eccentricity  
increases and tends towards unity. Hence, this changing orbit potentially   
permits collision with the host star or may cause the planet to leave  
the adiabatic regime.  
  
Although a planet is unlikely to reside in this specific  
configuration, it is useful to assess how nearby configurations evolve  
and how the planet's orbit is affected.  We first show that a planet  
will {\it eventually} achieve $e \rightarrow 1$ for all $a_0$ and  
$e_0$ values in the top two panels of Fig. \ref{fig:disctwo}, given  
$i(t) = i_0 = 90^{\circ}$ and $\omega(t) = \omega_0 =  
\arcsin{(\sqrt{1/5})}$.  The resulting survival timescale is higher  
for tighter planetary orbits and more circular orbits.  In the figure,  
we adopt $R = 4$ kpc, approximately halfway between the Earth and the  
Galactic centre. This choice is motivated by the microlensing planet  
searches, which are most sensitive to lenses half-way between observer  
and source star.  
  
Next, we consider small deviations from the stationary orbit.  We  
sample a wide variety of values of $\omega_0$, and show in the middle  
panels of Fig.~\ref{fig:disctwo} that for the initial values sampled,  
at $i=90^{\circ}$, $\omega_0$ asymptotically tends towards either  
$\omega_{\rm crit}$ or $(180^{\circ} + \omega_{\rm crit})$.  Doing so  
eventually causes the eccentricity to tend to unity.  Therefore, the  
value of $\omega_0$ does not appear to affect the final outcome, just  
the survival timescale.  
  
By contrast, even if $\omega_0 = \omega_{\rm crit}$, deviations from  
$i = 90^{\circ}$ will prevent the eccentricity from reaching unity,  
but still periodically increase its value to nearly unity.  The bottom  
two panels of Fig. \ref{fig:disctwo} illustrate this effect.  In those  
panels, the planets survive even if $i_0 = 85^{\circ}$.  However, the  
eccentricities of those planets achieve values so close to unity that  
they may become unstable due to other factors, such as close passage  
to the parent star or a small impulsive kick from other Galactic  
phenomena.  Note importantly from the bottom right panel that even the  
$i = 50^{\circ}$ planet, a full $40^{\circ}$ astride from the  
stationary inclination solution, exhibits eccentricity variations of  
several tenths.  Further, the bottom left panel demonstrates that,  
although $\omega$ is not stationary, it hovers around $\omega_{\rm  
  crit}$, $180^{\circ} - \omega_{\rm crit}$ or $180^{\circ} +  
\omega_{\rm crit}$ for the majority of the evolution.  
  
Returning to the strictly stationary orbit, we can estimate the survival  
timescale, $\tsurv$, analytically.  By Taylor expanding the  
inclination about $i = 90^{\circ}$ to first order and the eccentricity  
about $e=0$ to fourth order, we obtain an analytic solution to the  
first terms of Eqs. (\ref{adive}) and (\ref{adivi}).  The solution  
gives, for $\omega_0 = \omega_{\rm crit}$ and $\omega_0 = -\omega_{\rm  
  crit}$ respectively,  
\begin{equation}  
e(t) \approx  
\pm e_0 \frac{\sqrt{2}}  
{\sqrt{e_{0}^2 +  \left(2 - e_{0}^2\right) \exp{\left[2t \Upsilon_{zz}/n\right]}   }    }  
\end{equation}  
\begin{equation}  
e(t) \approx  
\pm e_0 \frac{\sqrt{2} \exp{\left[t \Upsilon_{zz}/n\right]}}  
{\sqrt{2 - e_{0}^2 + e_{0}^2\exp{\left[2t \Upsilon_{zz}/n\right]}   }    }  
\end{equation}  
In both cases, we obtain the same value of the survival timescale $\tsurv$  
when setting $e = 1$, namely  
\begin{equation}  
\tsurv \approx \left| \frac{n}{2\Upsilon_{zz}} \right|  
\ln{\left[ \frac{2 - e_{0}^2}{e_0^2}  \right]}  
\label{tsurv}  
\end{equation}  
Note that $\tsurv \approx n/\Upsilon_{zz}$ at $e_0 \approx 0.49$.  We  
overplot this analytical estimate with large dots on the upper panel  
of Fig. \ref{fig:disctwo} to demonstrate the quality of the  
approximation.  As expected, the approximation worsens as $e_0$ tends  
towards unity instead of zero.  
  
Equation (\ref{tsurv}) suggests that any planet on an adiabatic polar  
stationary orbit has a finite survival time.  For a main sequence  
lifetime $t_{ms}$, the critical planetary orbital period around its  
parent star for which a planet does not survive is  
\begin{equation}  
T_{\rm crit} = \frac{\pi}{t_{\rm ms} \Upsilon_{zz}} \ln{\left[ \frac{2  
      - e_{0}^2}{e_0^2} \right]}  
\end{equation}  
If we apply the disc density law from Eq. (\ref{disc}), then  
\begin{eqnarray}  
T_{\rm crit} &=& \frac{h}{2 t_{\rm ms} G \Sigma_0 } \ln{\left[ \frac{2 - e_{0}^2}{e_0^2}  \right]}  
\exp{\left(  R/ \Rd \right)}   
\\   
&\approx& 65400 {\rm yr} \times  
\ln{\left[ \frac{2 - e_{0}^2}{e_0^2}  \right]}   
\left( \frac{t_{\rm ms}}{10 {\rm Gyr}} \right)^{-1}  
\times  
\nonumber  
\\  
&&\left(\frac{h}{0.3 {\rm kpc}} \right)  
\exp{\left( \frac{R - 8 {\rm kpc}}{R_d } \right)}  
\end{eqnarray}  
For a star at $R = 4$ kpc, this yields a critical timescale on the  
order of $10^4 - 10^5$ yr.

\begin{figure*}  
\centerline{  
\psfig{figure=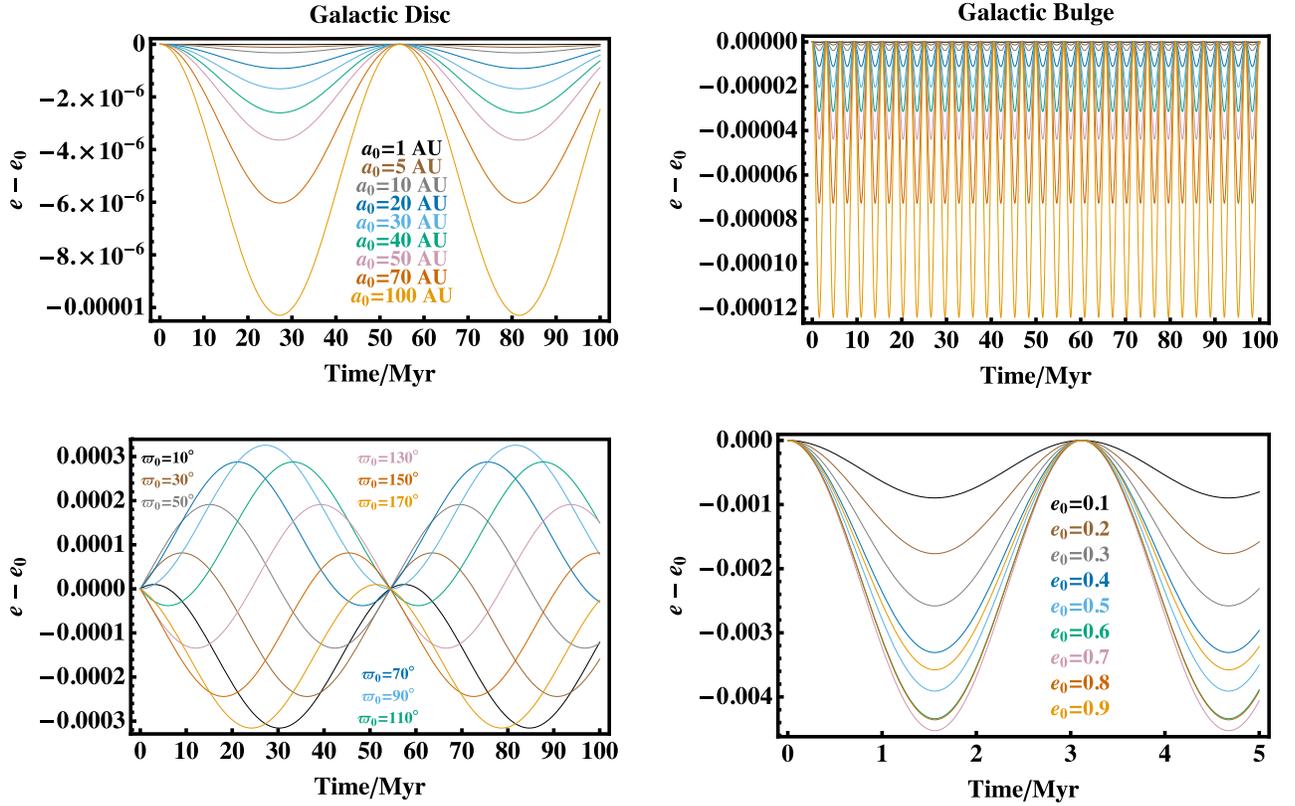,width=18cm}  
}  
\caption{The minimum possible eccentricity evolution for a selection  
  of exoplanets in the disc ($R = 4$ kpc) and bulge ($R = 0.2$ kpc)  
  regimes.  This limit is obtained when $i=0^{\circ}$; the evolution  
  is according to Eqs. (\ref{ppaasimprot})-(\ref{ppaomsimprot}).  In  
  the upper panels (with $e_0 = 0.5$), note the difference in both the  
  frequency and amplitude of the oscillations.  The initial semimajor
  axes for the curves in both upper panels are equivalent; in the lower
  panels, $a = 1000$ AU.  The lower-left panel  
  (also with $e_0 = 0.5$) demonstrates how the variations are  
  phase-shifted according to $\varpi_0$.  In the lower-right panel,  
  $\varpi_0 = 0^{\circ}$ is assumed for all curves, to demonstrate the  
  dependence on $e_0$; the amplitude peaks for $e_0 = 1/\sqrt{2}  
  \approx 0.7$.  }  
\label{fig:mine}  
\end{figure*}  

\section{The Galactic Bulge Regime} \label{sec:bulge}  
  
Exoplanet evolution in the Galactic bulge regime is more complex due  
to the importance of both the planar and vertical tides. The vertical  
tide is composed of the contribution from the bulge, but for locations  
in the outer bulge (such as at $R = 1$ kpc), we must also include the  
contribution from the disc.  Because of the interplay between the  
vertical and planar tides, attaining compact analytical results is  
difficult.  Hence, we perform numerical simulations to explore phase  
space.  
  
The results are summarized in Fig. \ref{fig:bulge1}.  All plots  
demonstrate a variety of eccentricity profiles; the upper plots model  
a wide-orbit planet ($a=10^3$ AU) and the lower plots illustrate the  
evolution for a closer-in planet (at $a=100$ AU and $30$ AU).  All  
plots assume $\varpi_0=0^{\circ}$, except the lower right plot, for  
which $R = 50$ pc and $e_0 = 0.5$ is assumed for all curves.  The  
difference in the evolution profiles from the upper plots (between  
the $R = 0.2$ kpc and $R = 0.3$ kpc curves) arises from the initial  
sign of $de/dt$, which is determined by the initial relative  
magnitudes of Eqs. (\ref{adive}) and (A2).  
  
Wide-orbit planets are significantly affected at all locations in the  
bulge.  In the innermost regions, the timescale of the eccentricity  
oscillations is fast: on the order of $1\%$ of the typical main  
sequence lifetime.  Also, the evolution in these innermost regions  
(within a couple hundred parsecs) becomes largely independent of  
inclination, as shown by the upper right plot.  However, other  
locations in the bulge are affected by changes of inclination.  The  
upper-right panel shows a similar evolutionary pathway to the  
upper-right panel of Fig. \ref{fig:disctwo}, even though here $\omega  
= 0^{\circ}$ and not $\omega_{\rm crit}$.  
  
Tighter-orbit bodies, such as analogues of the trans-Neptunian object  
Ceto ($a \approx 100$ AU) and the planet Neptune itself ($a \approx 30$ AU),   
are affected significantly only if  
the host stars reside within the inner hundred parsecs of the bulge.  
The lower-left plot illustrates the drastic differences in  
eccentricity profiles due to shifting the value of $R$ by just 150 pc.  
The lower-right plot presents the dependence on $\omega_0$ for an  
eccentric Neptune at $R = 50$ pc.  This planet cannot retain its  
primordial eccentricity.  If, however, the planet was born on a more  
circular orbit, the extent of the eccentricity change would decrease.  
  

\section{Minimum Exoplanet Eccentricity}  \label{sec:mine}  
  
A planetary system whose invariable plane varies little from the  
Galactic plane may be modeled in the planar adiabatic limit.  Assuming  
that the planet orbits the star in the same sense that the star orbits  
the Galactic centre (``prograde''), the equations of motion become:  
\begin{eqnarray}  
\frac{da}{dt}  
&=& 0  
\label{ppaasimprot}  
     \\  
\frac{de}{dt}  
&=&   
\frac{5e\Omega_{G}^2\sqrt{1-e^2}}{2n}  
\sin{\left[ 2 \left(\varpi - \Omega_{G} t \right) \right]}  
\label{ppaesimprot}  
    \\  
\frac{di}{dt}  
&=& 0     
     \\  
\frac{d\varpi}{dt}  
&=&     
\frac{5\Omega_{G}^2\sqrt{1-e^2}}{2n}  
\cos{\left[ 2 \left(\varpi - \Omega_{G} t \right) \right]}  
\label{ppaomsimprot}  
\end{eqnarray}  
  
We plot solutions of these equations in Fig. \ref{fig:mine} for  
planets in the disc at $R = 4$ kpc (left panels) and in the bulge at  
$R = 0.2$ kpc (right panels).  We select $e_0 = 0.5$ for all plots  
except the lower-right plot.  The upper panels show that planets in  
the disc regime typically feature eccentricity variations of $10^{-6}$  
to $10^{-5}$ with a period of tens of Myr; planets in the bulge regime  
show variations of $10^{-5}$ to $10^{-4}$ with a period of a few Myr.  
These panels illustrate only eccentricity decreases because $\varpi_0  
= 0^{\circ}$ for each of those curves.  We sample $\varpi_0$ at nine  
equally spaced values from and including $10^{\circ}$ and  
$170^{\circ}$ in the lower-left panel, demonstrating the sensitive  
dependence of the eccentricity evolution on $\varpi_0$.  This  
dependence is equivalent in the disc and bulge.  The lower-right panel  
instead illustrates the dependence of amplitude on $e_0$ (for  
$\varpi_0 = 0^{\circ}$).  This relationship between amplitude and  
$e_0$ is not monotonic because of the $e \sqrt{1-e^2}$ term in  
Eq. (\ref{ppaesimprot}).  Consequently, the greatest variation occurs  
for $e_0 = 1/\sqrt{2} \approx 0.7$.  
  
This term prevents a complete analytical solution to Eqs.  
(\ref{ppaasimprot})-(\ref{ppaomsimprot}).  However, under the small  
eccentricity approximation, one may attain a closed analytical  
solution for the eccentricity evolution.  The solution gives the  
following maximum eccentricity increase and decrease factors:  
\begin{equation}  
\approx         
1 \pm \frac{5\Omega_{G}}{2n}  
\approx  
1 \pm   
\frac{9 \times 10^{-8} T}{R} {\rm kpc}/{\rm yr}^2   
\label{base1}  
\end{equation}   
where $T = 2\pi/n$ is the planet's orbital period about its parent  
star and we used the approximations in Eqs. (\ref{AB})-(\ref{Oort}).  
Therefore, all exoplanets in the galactic disc have eccentricities  
which vary by a factor of at least $(1 \pm 5\Omega_{G}/2n)$.  Note  
that one cannot use Eq. {(\ref{base1})} to determine when a planet  
will survive because this estimate is based on the adiabatic  
approximation and for low eccentricity.  
  
Now we consider how Galactic tides affect the pericentre advance or retreat of  
planets.  The maximum variation of $d\varpi/dt$ is equal to  
\begin{equation}  
\frac{5\Omega_{G}^2}{2n} =  
\frac{2 \times 10^{-14} T}{R^2} {\rm rad} \times {\rm kpc}^2/{\rm yr}^2  
\end{equation}  
If we compute the critical semimajor axis at which the maximum  
pericentre precession rate from Galactic tides is comparable to   
that from general relativity, we obtain:  
\begin{eqnarray}  
a_{\rm crit} &=& \left( \frac{6}{5} \right)^{\frac{1}{4}}   
\sqrt{ \frac{G \left(m_{\star} + m_p \right)}{c \Omega_{G}}}  
\left(1 - e^2\right)^{-\frac{1}{4}}  
\nonumber  
\\  
&\approx&  
55 {\rm AU} \left(\frac{R}{1 {\rm kpc}}\right)^{\frac{1}{2}}   
\left(\frac{m_{\star}}{M_{\odot}}\right)^{\frac{1}{2}}  
\left(1 - e^2\right)^{-\frac{1}{4}}   
\end{eqnarray}  
where $c$ is the speed of light.  We emphasize that these precession   
rate estimates are lower bounds because they were derived in the   
limiting case of planar adiabatic motion.

\section{Discussion} \label{discussion}

\subsection{Implications from Tides}

Here, we discuss some of the implications of the tidal results that we
have obtained from this work.  First, the inclination of the planetary
orbit with respect to the Galactic plane might be indicative of
dynamical excitation in a planetary system, and vice-versa.
Particularly, planetary orbits that are highly inclined to the
Galactic plane will feature the greatest excitation.  Although
survival is likely for the smallest orbits in these systems, their
eccentricity variations at a given (fixed) semimajor axis will be
higher than in other systems.  Despite this variation periodically
becoming zero, such periods of dynamical quiescence represent
typically just a small fraction of the parent star's main sequence
lifetime.  Our lack of unbiased exoplanetary inclination data suggests
that we cannot yet pinpoint a preferential planetary inclination with
respect to the Galactic disc.  However, our own Solar System and the
variety of orientations exhibited by transiting planets prove that $i
\ge 40^{\circ}$ can easily exist for planetary systems.
  
Second, our results are not restricted to planets.  Our analysis may
be extended to binary stars or belts of objects such as Kuiper belts
or scattered discs.  \cite{jiatre2010} study the evolution of wide
binary stars in the Solar neighbourhood.  One can instead consider
this evolution at other locations in the Galaxy by setting $m_p =
m_{\star}$ in our Eqs. (\ref{xeq})-(\ref{zeq}).  For binary stars of
equal masses, $n$ will be increased by a factor of $2^{1/2}$, and
hence, the time evolution of every variable
(Eqs. \ref{adiva}-\ref{adivvarpi} and \ref{adia}-\ref{adOi}) will be
decreased by a factor of 1.41.  The amplitudes of the variations will
otherwise remain unaffected.  Modeling Kuiper belts and scattered
discs involves imposing distributions of initial conditions on our
equations.  Because the individual bodies in these belts are unlikely
to interact with one another, they can be treated by our formalism.
Oort clouds are typically too distant to be treated in the adiabatic
approximation.  Their evolution must be modelled either with N-body
simulations \citep{Br01,kaietal2011} or with the nonadiabatic
equations of motion \citep{Ve12}.
  
Third, differential pericentre precession due to Galactic tides might  
affect long-term N-body simulations of planetary systems.  \cite{Ve10}  
highlighted the danger of neglecting general relativity when modeling  
hierarchical multi-body systems with high relative inclinations, and  
the effect from Galactic tides might equally be important to  
incorporate, particularly at high inclinations with respect to the  
Galactic disc.   
  
Fourth, improving observational precision might be able to place
meaningful constraints on planetary systems.  Currently, the smallest
observational errors on planetary eccentricity measurements are about
$1 \times 10^{-3}$ \citep{Wo94,We12}.  Although these values are well
above the baseline eccentricity variations predicted by
Eq. (\ref{base1}), and are measured for tight orbits, they might be
comparable to expected eccentricity variations for planet WD 0806-661,
with $a \approx 2500$ AU \citep{Lu11}.  Two other planets with $a >
1000$ AU \citep[e.g.][]{Go10,Ku11} bolster theories that planets can
exist in such remote regions of planetary systems despite being formed
elsewhere.

Fifth, the influence of Galactic tides may represent another way in 
which planets can transform into Hot Jupiters.  Near-polar
planetary orbits might be highly eccentric 
(see Fig. \ref{fig:disctwo}).  The pericentres of these
planets might reside close enough to the star such that 
{\it star-planet} tides will damp both the semimajor axis and
eccentricity of the planet.  The frequency of such transformed
Hot Jupiters would be an increasing function of both age and 
Galactocentric distance.  This mechanism is unlikely to be
prevalent in the Solar Neighbourhood because the Galactic 
tidal timescales are likely to be too long to allow for
star-planet tides to shrink the planetary orbits appreciably.

\subsection{Effects of Passing Stars}

Although stars collectively help establish the Galactic tide,
individually they can produce strong perturbations when
they fly past a planetary system.  The consequences of this brief
perturbation may be comparable to tides acting over Gyr.
The effect of flybys on planetary systems has been studied 
before in the Galactic Disc \citep[e.g.][]{zaktre2004,vermoe2012}.
Here we quantify the effect in the Galactic bulge.

We use the formalism of \cite{vermoe2012}, which is sufficiently
general to be applied to the bulge.  They assumed every star has
a single planet on an initially circular orbit, and computed
cross sections for the planetary eccentricities to be perturbed
by a given amount, $\kappa$.  We simplify their treatment by
assuming each planet has a mass of $1M_J$ and initially 
resides at the same distance from their parent stars, $a_{\rm same}$,
all of which are have masses of $1M_{\odot}$.  Then the number
of times, $\mathcal{N}$, which a planet's eccentricity will
be changed by $\kappa$ over a star's main sequence lifetime is equal to:

\begin{eqnarray}
\mathcal{N} &=& 0.0905 \eta \left(\frac{a_{\rm same}}{1000 \ {\rm AU}}\right)^{\frac{3}{2}} \left(\frac{\rho_{\rm bulge}}{0.5 M_{\odot}{\rm pc}^{-3}}  \right)
        \left(\frac{t_{\rm ms}}{10^{10}{\rm yr} }  \right)
\nonumber
\\
&\times&
\sigma_{\rm norm}\left(\left|\Delta e \right|>\kappa, \eta \right)
\end{eqnarray}

\noindent{where} the normalized cross sections $\sigma_{\rm norm}$
can be read off directly from Figs. 8-11 of \cite{vermoe2012},
and $\eta \equiv V_{\rm dispersion}/V_{\rm critical}$.  We adopt
a dispersion velocity of $200$ km/s as representative of
three-dimensional bulge dispersion velocities.  The critical velocity
is the encounter velocity at which the total energy of the system
is zero and ionization is possible \citep[e.g.][]{freetal2004}:

\begin{equation}
V_{\rm critical} =
2\sqrt{ 
\frac{G M_{\odot} M_J}
{a_{\rm same} \left(M_{\odot} + M_J \right)}  
     }
\end{equation}

\begin{figure}  
\centerline{  
\psfig{figure=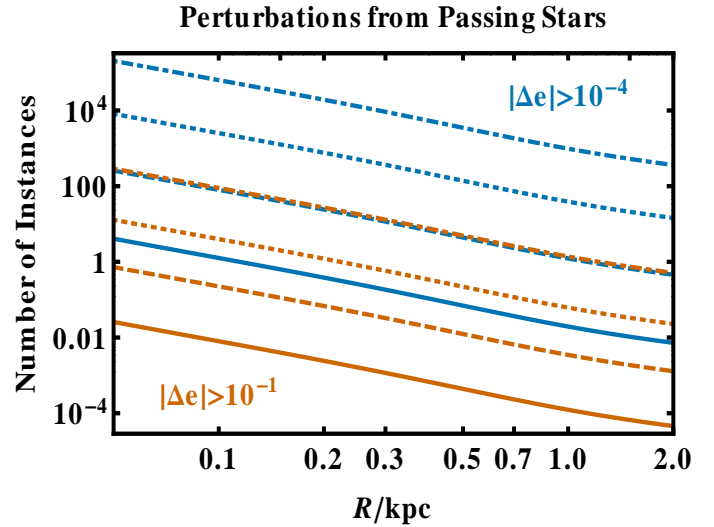,width=9.5cm,height=7cm}  
}  
\caption{Eccentricity changes from passing stars in the bulge.  Shown
  are the number of instances over a main-sequence lifetime that a
  planet's eccentricity should change by $10^{-1}$ (orange lines) or
  $10^{-4}$ (blue lines).  Each set of four lines, from bottom to top
  (solid, dashed, dotted, dot-dashed) correspond to $a_{\rm same} =
  \left\lbrace 1, 10, 100, 1000 \right\rbrace$ AU.  The plot shows
  that the eccentricity changes from stellar flybys in the bulge may
  be comparable to changes induced by the Galactic tides.  }
\label{fig:bulge}  
\end{figure}

We consider $\kappa = \left\lbrace 10^{-4}, 10^{-1} \right\rbrace$,
$a_{\rm same} = \left\lbrace 1, 10, 100, 1000 \right\rbrace$ AU and
compute $t_{ms} = 10.941$ Gyr by assuming stellar metallicity from
\cite{huretal2000}.  Figure \ref{fig:bulge} demonstrates that the
minimum possible eccentricity variations due to tides presented in
Fig. \ref{fig:mine} is likely to be surpassed by passing stars.  In
the opposite extreme, the eccentricity evolution of planets on wide
polar orbits appears to have comparable contributions from tides and
stellar flybys (see Fig. \ref{fig:bulge1}).  Therefore, both of these
effects may be equally important and deserve detailed study.

\section{Conclusions} \label{conclusion}  
  
The effects of Galactic tides have been studied before in the Solar  
neighbourhood. This is the first investigation into the effect of  
tides on planetary systems within the Oort cloud throughout the rest 
of the disc of our Galaxy, as well as in the bulge.  
  
In the Solar neighbourhood, the Hill surface has an extent of $\sim
10^5 (M/M_\odot)^{1/3}$ AU. For the Sun, this marks the outer boundary
of the Oort Cloud, and Galactic tides are known to play an important
role in dislodging comets into the inner Solar
System~\citep{smotor1984,He86,dunetal1987,Ma89,levetal2001,kaiqui2009}.
However, the effect of Galactic tides on the Sun's planets is
negligible, even over the age of the Solar system.  However, the
architecture of exoplanets is very varied, and wide-orbit exoplanets
with semimajor axes between 100-2500 AU are known to exist~\citep[see
  e.g.,][]{kaletal2005,Go10,Lu11, Ku11}. In the inner Galaxy, the
tidal forces are stronger, typically scaling inversely with radius in
the Galaxy. For example, in the inner 500 pc, the Hill surface has a
size of $\sim 10^4 (M/M_\odot)^{1/3}$ AU, whilst the timescale over
which evolutionary effects become noticeable is billions of years.
  
The Galactic tide is dominated by the contribution from the Galactic
disc over radii in the range 3.5 to 10 kpc. In this disc regime, the
popular practice of neglecting the planar tides is justifiable.
Planetary systems that are at least moderately inclined to the
Galactic disc are susceptible to slow but significant planetary
eccentricity evolution over the parent star's main sequence lifetime.
At high inclinations to the Galactic plane, eccentricity evolution is
greater. As the semimajor axis is never secularly affected by tidal
perturbations, the periastron tends to zero and the planet is driven
towards the host star on timescales of $\gtrsim 10^9$ yr.
  
Within 3.5 kpc of the Galactic Centre, the contribution of the
Galactic bulge is always important. As the matter distribution is more
spherical, the planar tidal components are as important as the
vertical ones. Evolution in the bulge regime is now a result of
complex interplay between the vertical and planar tides. Wide orbit
planets ($a = 1000$ AU) are substantially affected by tides on
timescales (tens to hundreds of Myr) that are much smaller than the
main sequence lifetime ($\approx 10$ Gyr). Eccentricity variations of
several tenths are very typical for these planets.  They are never in
a state of quiescence; their orbital parameters are continually
changing though the effects of tides. Closer-in planets, at e.g., 100
AU, are only affected if the host star resides within the inner
hundred parsecs.
  
Our study of tides is a first step in the understanding of how  
exoplanetary systems interact with their Galactic environment. The  
main limitation of our work is that the host star has been assumed to  
move in a circular orbit in the Galactic plane. In fact, stars lead  
much more exciting lives! They are usually inclined to the Galactic  
plane, their orbits are usually eccentric and sometimes chaotic, and  
they suffer perturbations that can move them many kiloparsecs. There  
is, for example, both chemical and dynamical evidence that the Sun has  
been moved substantially from its place of birth~\citep{Cl97, Se02} by  
spiral waves. There are therefore good reasons for believing that the  
effects of tides are still more substantial than we have found here!  
  
\section*{Acknowledgments}  

We thank an anonymous referee for a helpful and probing report, and 
Fred C. Adams for reading through the manuscript and providing detailed, 
useful feedback.

\appendix  
\onecolumn  
  
\section{Planar Adiabatic Equations}  
  
\  
  
We obtain the planar adiabatic equations for a planet subjected to tidal perturbations $\Upsilon$ 
by setting $\Upsilon_{xy} = \Upsilon_{yx}$ in Eqs. (25)-(29) of \cite{Ve12}:
\begin{eqnarray}  
\left(\frac{da}{dt}\right)_p  
&=& 0  
\label{adia}  
\\  
\left(\frac{de}{dt}\right)_p  
&=&  
\frac{5e \sqrt{1 - e^2}}{16 n} \bigg\lbrace  
\left[  
4 \cos{i} \cos{2\omega} \sin{2\Omega}   
+ \sin{2\omega}\cos{2\Omega} \left(3 + \cos{2i}\right)  
\right]  
(\Upsilon_{xx}+\Upsilon_{yy}) + 2 \sin^2{i}(\Upsilon_{xx} -\Upsilon_{yy} )  
\nonumber  
\\  
&-&  
2\left[  
4 \cos{i}\cos{2\omega} \cos{2\Omega}  
-  
\sin{2\omega}\sin{2\Omega} \left(3 + \cos{2i} \right)  
\right]  
\Upsilon_{xy}  
\bigg\rbrace  
\label{adie}  
\\  
\left(\frac{di}{dt}\right)_p  
&=&    
\frac{\sin{i}}{8n\sqrt{1 - e^2}}  
\bigg\lbrace  
\left[ \sin{2\Omega} \left(2 + 3 e^2 + 5 e^2 \cos{2 \omega} \right)\right](\Upsilon_{xx}-\Upsilon_{yy})  
-10 e^2 \cos{i} \sin{2\omega} \sin^2{\Omega} (\Upsilon_{xx}+\Upsilon_{yy})  
\nonumber\\  
&+& 20 e^2 \cos{i} \sin{2\omega} \cos{\Omega} \sin{\Omega}  
\Upsilon_{xy}\bigg\rbrace \label{adii}  
\\  
\left(\frac{d\Omega}{dt}\right)_p  
&=&   
\frac{1}{4n\sqrt{1 - e^2}} \bigg\lbrace  
\left[ \sin^2{\Omega} \cos{i} \left(-2 - 3 e^2 + 5 e^2 \cos{2 \omega} \right) \right](\Upsilon_{xx}+ \Upsilon_{yy})  
+ 5 e^2 \cos{\Omega} \sin{\Omega} \sin{2\omega} (\Upsilon_{xx} - \Upsilon_{yy})  
\nonumber
\\
&-& 
2\left[ \cos{\Omega} \sin{\Omega} \cos{i} \left(-2 - 3 e^2 + 5 e^2 \cos{2 \omega} \right)\right]\Upsilon_{xy} \bigg\rbrace
\label{adOi}
\\
\left(\frac{d\omega}{dt}\right)_p
&=&
\frac{1}{16n\sqrt{1-e^2}} \bigg\lbrace C_{12} (\Upsilon_{xx} + \Upsilon_{yy})
+
\left(2C_{10} + C_{11}\right) (\Upsilon_{xx} -\Upsilon_{yy})
-
2\left(2C_{10}\cot{2\Omega} + C_{13}\right) \Upsilon_{xy} 
\bigg\rbrace
\label{adoi}
\end{eqnarray}
where, using the notation of \cite{Ve12}, we have
\begin{eqnarray}
C_{10} &\equiv& 5\left(e^2-2 \right) \sin{(2\omega)} \sin{(2\Omega)} \cos{i}
\\
C_{11} &\equiv& \cos{2\Omega} \left(1 - 6e^2 - 5 
\left(-3 + 2 e^2 \right) \cos{2\omega} - 10 \cos{2i} \sin^2{\omega}  \right)
\\
C_{12} &\equiv& 11 - 6e^2 + \cos{2\omega} \left(5 - 10e^2 \right)
+
10 \cos{2i}\sin^2{\omega}
\\ 
C_{13} &\equiv& \sin{2\Omega} 
\left(-1 + 6e^2 + 5\left(-3 + 2e^2\right) \cos{2\omega}
+
10 \cos{2i} \sin^2{\omega} \right)
\end{eqnarray}

\label{lastpage}

\end{document}